\documentclass[draft, preprintnumbers, preprint, nofootinbib,
               noshowpacs, a4paper]
              {revtex4}

\usepackage{amsmath, amssymb}

\newcommand{\bal}{\begin{equation}\begin{aligned}}
\newcommand{\eal}{\end{aligned}\end{equation}}

\begin{document}
\title{Classification of Cohomogeneity One Strings}

\author{Hideki Ishihara}
\email{ishihara@sci.osaka-cu.ac.jp}

\author{Hiroshi Kozaki}
\email{furusaki@sci.osaka-cu.ac.jp}

\affiliation{
Department of Mathematics and Physics,
Graduate School of Science, Osaka City University,
Sugimoto, Sumiyoshi-ku, Osaka, 558-8585, Japan
}

\preprint{OCU-PHYS 231}
\preprint{AP-GR 25}


\date{\today}
\begin{abstract}
We define the cohomogeneity one string, string with  continuous symmetries, 
as its world surface 
is tangent to a Killing vector field of a target space. 
We classify the Killing vector fields by an equivalence relation 
using isometries of the target space.
We find that the equivalence classes of Killing vectors in Minkowski 
spacetime are partitioned into seven families. 
It is clarified that there exist seven types of  
strings with spacelike symmetries and four types of strings with 
timelike symmetries, stationary strings.
\end{abstract}

\maketitle

\section{Introduction}

Recently, extended objects, such as strings and membranes, 
gather much attention in the contexts of unified theories and cosmologies. 
A trajectory of the extended object, world hypersurface, is a 
submanifold embedded in a target space. 
The geometry induced on the world hypersurface has indeed much variety 
which depends on the possible solutions to the equations of motion 
of the object, but extended objects which have geometrical symmetries 
are especially interesting for their simplicity.


A matter of main interest in the brane universe scenario is 
the structure of symmetric extended objects, i.e.,  
spatially homogeneous brane universe models 
embedded in a symmetric bulk space\cite{brane}.
Another example is classical stationary solutions for 
Nambu-Goto string in stationary spacetimes
\cite{Frolov:1988zn,Carter:1989bs,Frolov:1996xw,Vilenkin-Shellard}.  
Equations of motion of an extended object governed by 
the Nambu-Goto action or its generalization 
has the form of partial differential equations with non-linear constraint 
equations in general. 
However, the equations reduce to ordinary differential equations 
when the object has enough symmetries. 
In the investigations noted above, rich properties of the extended 
objects are clarified by solving the ordinary differential equations 
explicitly. 

Suppose that an $m$-dimensional hypersurface is embedded in an 
$n (>m)$-dimensional target space which has isometries, 
the embedded hypersurface can 
possess a part of the symmetries of the target space. 
If a subgroup of the isometries of target space acts on the hypersurface,
 the restriction on the hypersurface is isometries of it. 
In addition, the extrinsic geometry of the hypersurface is also symmetric 
with the isometries. 
The hypersurface is said to be a cohomogeneity one hypersurface
if a subgroup of isometries of a target space 
acts on the hypersurface, and the orbits of the action on it span an
$(m-1)$-dimensional space. 
As is discussed later in the Nambu-Goto string case, 
the equations of motion for a cohomogeneity one 
object can be reduced to ordinary differential equations.

Among a variety of extended object's dimension,  
we concentrate ourself especially on cohomogeneity one strings, 
as the simplest extended objects, in this letter. 
The one-parameter group of isometries acting on the world surface is
generated by a Killing vector field on a target space, that is, 
the string's world surface is tangent to the Killing vector field.

If a target space has only one Killing vector, 
there is one class of cohomogeneity one strings 
associated with the Killing vector.
On the other hand, in a space with more than two independent Killing
vectors, there would be infinite classes of cohomogeneity one strings 
since infinite numbers of linear combinations of the Killing
vectors are possible.

Let us consider the four-dimensional Minkowski spacetime,
which has ten linearly independent Killing vectors, 
as a target space. 
Suppose two Killing vectors which denote translation symmetries along 
different directions. 
Though these Killing vectors are linearly independent, 
cohomogeneity one strings associated with these Killing vectors are 
members of the same class 
because the world surfaces of these strings are transformed each other 
by a suitable isometry of Minkowski spacetime which rotates one direction 
of translation symmetry into the other.

The issue treated in this letter is how many classes of cohomogeneity one
strings are there in a symmetric spacetime. 
For clarifying this problem, we classify the Killing vectors 
in Minkowski spacetime, as a typical example of symmetric spacetime, 
introducing an equivalence relation 
by using isometries of the spacetime.

\section{Equations of Motion of Cohomogeneity One String}

\newcommand{\M}{\cal M}
\newcommand{\N}{\cal N}
\newcommand{\C}{\cal C}

Let us consider a two-dimensional world surface $\Sigma$ is embedded in 
an $n$-dimensional target spacetime $\M$ with a metric $g_{ab}$ 
which possesses  Killing vector fields.
If the world surface $\Sigma$ is tangent 
to one of the Killing vector fields of $\M$, say $\xi^a$, 
we call the world surface $\Sigma$ a cohomogeneity one string associated 
with the Killing vector $\xi^a$. 
The stationary string is one of the example associated 
with a Killing vector which is timelike on the surface\cite{Frolov:1996xw}.

When we choose a Killing vector field in $\M$, 
there would be many cohomogeneity one world surfaces 
associated with the Killing vector. 
If the equations of motion governing the dynamics of the 
string is given, the possible cohomogeneity one string 
surfaces are selected out as solutions to the equations.
Hereafter, we consider the Nambu-Goto string, as the simplest example, 
whose action is given by
\bal
	S=\int_\Sigma \sqrt{-\gamma}, 
\label{NG-action}
\eal
where $\gamma$ is the determinant of induced metric on the world surface.

The orbit space of a Killing vector field $\xi^a$ of $\M$, say $\N$, 
is an $(n-1)$-dimensional space on which the metric
\bal
	h_{ab}=g_{ab}-\xi_a\xi_b/f
\label{hab}
\eal
is introduced naturally, where $f:=\xi^a\xi_a$.
The metric $h_{ab}$ has Euclidean signature in the region where $f<0$ 
and Lorentzian signature in $f>0$.

Suppose that a curve $\C$ is given in $\N$, 
orbits of the Killing vectors $\xi^a$ starting from $\C$ span 
a two-dimensional surface. 
Since $h_{ab}$ measures the length along the direction perpendicular to 
$\xi^a$, area of the surface element is simply given by
\bal
	dA = \sqrt{|f|}\sqrt{|h_{ab}dx^adx^b|}, 
\eal
where $dx^a$ is an infinitesimal displacement along $\C$.
Therefore, the action \eqref{NG-action} reduces to 
\bal
	S=\int_{\C} \sqrt{-\tilde h_{ab}dx^adx^b}, 
\label{line_action}
\eal
where 
\bal
	\tilde h_{ab}:= f h_{ab}. 
\label{thab}
\eal
The action \eqref{line_action} gives the length of $\C$ 
with respect to the metric $\tilde h_{ab}$ on $\N$. 
Therefore, the problem for finding solutions of cohomogeneity one string 
associated with $\xi^a$ reduces to 
the problem for solving $(n-1)$-dimensional geodesic equations 
with respect to the metric $\tilde h_{ab}$.
When a world surface is associated with a spacelike Killing vector, i.e., 
$f>0$ on the surface, $\C$ should be a timelike geodesic 
with respect to the metric $\tilde h_{ab}$ so that  
$\Sigma$ is a timelike world surface.

\section{Classification in Minkowski Spacetime} 
\newcommand{\bs}{\boldsymbol}
\newcommand{\bp}{\boxplus}

First, let us consider a pair of two-dimensional world surfaces, 
$\Sigma_{A}$ and $\Sigma_{B}$, embedded 
in a symmetric target space $\M$ which admits isometries. 
The world surfaces $\Sigma_{A}$ and $\Sigma_{B}$ 
are geometrically equivalent if there is an isometry 
$\phi$ of $\M$ which maps $\Sigma_{A}$
onto $\Sigma_{B}$: 
\begin{align}
 \phi: \Sigma_{A} \rightarrow \Sigma_{B}.
\end{align}

Next, let $\Sigma_A$ be a cohomogeneity one string associated with a 
Killing vector field $\bs\xi_A$ and let 
$\Sigma_B$ be a geometrically equivalent to $\Sigma_A$. 
The isometry $\phi$ which maps $\Sigma_A$ 
to $\Sigma_B$ pushes forward $\bs\xi_A$ to a vector field, say $\bs\xi_B$: 
\begin{align}
 \phi^{\ast}: \bs\xi_{A} \rightarrow \bs\xi_{B}.
\end{align}
It is clear that $\bs\xi_B$ is also a Killing vector field on $\M$, 
and $\Sigma_B$ is a cohomogeneity one string associated with it. 
If $\M$ has a large number of symmetries, it is possible that 
$\bs\xi_A$ and $\bs\xi_B$ are linearly independent vector fields 
which are connected by the isometry $\phi$. 
Therefore, it suggests the idea that we classify Killing vector fields 
by the equivalence relation with respect to the isometries.

We introduce an equivalence relation, $\sim$, as follows: 
Killing vectors $\bs\xi_{A}$ and $\bs\xi_{B}$ are equivalent 
if and only if there is an isometry $\phi$ which pushes
forward $\bs\xi_{A}$ to $\lambda \bs\xi_{B} ~(\lambda :\text{constant})$: 
\begin{align}
	\phi^{\ast}: \bs\xi_{A} \rightarrow \lambda \bs\xi_{B} \quad 
	\Leftrightarrow \quad \bs\xi_{A} \sim \bs\xi_{B}. 
\label{equiv}
\end{align}
The scalar multiplication comes from the fact that the Killing 
vector field is irrelevant to the constant scaling.

Here, we classify the Killing vectors in Minkowski spacetime based on 
the equivalence relation noted above.
In Minkowski spacetime with the metric 
\bal
	ds^2 = - dt^2  + dx^2 + dy^2 + dz^2, 
\label{Minkowski}
\eal
there are ten independent Killing vectors which generate Poincar\'e group: 
\begin{align*}
	&\bs{P}_{a}~(a=t, x, y, z), 
	&&\mbox{Translations along $a$-direction;}&
\\
	&\bs{K}_{i}~(i = x, y, z),
	&&\mbox{Lorentz boosts along $i$-direction;} &
\\
	&\bs{L}_{i}~(i = x, y, z),
	&&\mbox{Space rotations around $i$-axis.} &
\end{align*}
An arbitrary Killing vector $\bs{\xi}$ is expressed as
a linear combination of them: 
\begin{align}
 \bs{\xi}
 	=  \alpha_{a} \bs{P}_{a}
   		+ \beta_{i}  \bs{K}_{i} + \gamma_{i} \bs{L}_{i},
 \label{eq:arbitrary_killing}
\end{align}
where $\alpha_{a}$, $\beta_{i}$ and $\gamma_{i}$ are constant coefficients.

We will list up inequivalent representatives under the equivalence 
relation \eqref{equiv} using Poincar\'e group of isometries. 
The task we should do is logically straightforward but tedious. 
Then, we would like to show some typical procedure and 
final result.

As an example of the procedure of classification, 
we consider a Killing vector in the form
\bal
	\bs\xi = a \bs{K}_y + b \bs{L}_z,
\label{axi}
\eal
where $a$ and $b$ are arbitrary constants. 
We use the Lorentz boost along $x$-direction 
denoted by 
$\phi=\textrm{Exp}[\varphi \bs{K}_x]$, where $\varphi$ is 
a parameter.
The Killing vector
$\bs\xi$ 
is pushed forward by $\phi$ as
\begin{align}
	\phi^*: \bs\xi
 	\rightarrow 
  	(a \cosh \varphi - b \sinh \varphi) \bs{K}_y
 	+ (-a \sinh \varphi + b \cosh \varphi)\bs{L}_z.
\end{align}
Choosing the parameter $\varphi$ suitable for 
the values of $a$ and $b$, 
we see that $\bs\xi$ is equivalent to one of three Killing vectors:
\begin{align}
	\bs\xi= &a \bs{K}_y + b \bs{L}_z 
 \sim
 \begin{cases}
  \bs{K}_y &\text{for $|a| > |b|$}\\
  \bs{L}_z &\text{for $|a| < |b|$}\\
  \bs{K}_y + \bs{L}_z 
  & \text{for $|a| = |b|$}
 \end{cases} , 
\label{eq:example1}
\end{align}
where we also use space or time reflection.

In contrast, it is worth while to note that the Killing vector in the form
\bal
	\bs\xi=a\bs K_z + b\bs{L}_z,
\eal
for example, is not equivalent to 
\bal
	\bs\xi'=a'\bs K_z + b'\bs{L}_z
\eal
except the case $|a/b|= |a'/b'|$. 
In the metric \eqref{Minkowski}, the norm of $\bs\xi$ is given by
\bal
 	a^2(-z^2+t^2)+b^2(x^2+y^2), 
\eal
Then, equinorm surfaces of $\bs\xi$ are in the form of squashed hyperboloids, 
where $a^2/b^2$ describes the amount of squashing. 
Since the pushing forward $\phi^*$ preserves 
the norm of vectors then the shape of the equinorm surfaces should be same 
for equivalent vector fields. 
If $|a/b|\neq |a'/b'|$ the amount of squashing of 
the hyperboloid is different. 
Thus, there is no isometry which maps $\bs\xi$ to $\bs\xi'$, i.e., 
$\bs\xi$ is inequivalent to $\bs\xi'$.

Starting from a general linear combination
(\ref{eq:arbitrary_killing}), 
we can complete the classification. 
We find that equivalence classes are partitioned into 
seven families listed in Table I. 
Each partition consists of infinite classes parameterized by one parameter. 
We introduce a notation, $\bp$, such that  
$\bs{P}_{t}\bp \bs{L}_z$, for example, means the set of all 
linear combinations $a\bs P_t + b\bs{L}_z$,  
where $a$ and $b$ are non-negative constants satisfying $a^2+b^2=1$. 
The constraints on $a$ and $b$ are for eliminating total 
scaling of Killing vectors and redundancy under space and time reflections.

\setlength{\tabcolsep}{10pt}
\begin{table}
\begin{tabular}{cccc}
\hline
 &\mbox{family of classes}&\mbox{equinorm surface}&
\mbox{timelike region}\cr
\hline
I &$ \bs{P}_{t}\bp \bs{L}_z$
	&$-a^2+b^2(x^2+y^2)=const. $& \mbox{yes} 
\cr
II &$(\bs{P}_{t} + \bs{P}_{z})\bp \bs{L}_z $
	&$ b^2(x^2+y^2)=const. $& \mbox{no}
\cr
III &$\bs{P}_z\bp\bs{L}_z $
	&$a^2+b^2(x^2+y^2)=const.$& \mbox{no}
\cr
IV &$\bs{P}_z \bp(\bs{K}_{y} + \bs{L}_z)$ 
	&$a^2+b^2(t+x)^2=const. $&  \mbox{no}
\cr
V &$\bs{P}_z\bp\bs{K}_y$ 
	&$a^2+b^2(t^2-y^2)=const.$& \mbox{yes}
\cr
VI &$\bs{P}_x\bp (\bs{K}_{y} + \bs{L}_z)$
	&$a^2-2ab y+b^2(t+x)^2=const. $& \mbox{yes}
\cr
VII &$\bs{K}_z\bp\bs{L}_z$
	&$ a^2(t^2-z^2)+b^2(x^2+y^2)=const.$&  \mbox{yes} \cr
\hline
\end{tabular}
\caption{ 
Families of equivalence classes. 
Equinorm surfaces of the representatives in Minkowski spacetime, 
and whether timelike region of Killing vectors
exists or not are shown. 
}
\end{table}
%

The number of partitions would crucially depends on the 
structure of the group of isometries. Since Poincar\'e group 
is the group of isometries preserving indefinite Minkowski metric then 
the first three families, I$\sim$ III, should be distinguished. 
If the classification is done in Euclidean space, these 
families would fall into one family since the rotations in $t-z$ plane 
connect members in these families. 
Similarly, families of III, IV and V 
would fall into one family in the flat Euclidean space.
In addition, it seems that the existence of translation group 
makes the issue complicated. It would be interesting to classify 
the Killing vectors in other maximally symmetric spacetime: 
de Sitter and anti-de Sitter spacetimes\cite{desitter}. 

The Killing vector fields in the families 
I, V, VI and VII can be timelike in 
some regions in Minkowski spacetime except edges of the families 
i.e., $\bs L_z, \bs P_z $, and $\bs K_y + \bs L_z$. 
Then, there are four types of stationary strings 
which correspond to four families of timelike Killing vectors. 
The rigidly rotating strings are one of them; 
they are associated with the Killing vectors in 
the family $\bs{P}_{t}\bp\bs{L}_{z}$.
All Killing vector fields except $\bs P_t$ and $(\bs P_t+ \bs P_z)$ 
have spacelike region then there are seven types of cohomogeneity one 
strings with spacelike symmetries. 

From \eqref{hab} and \eqref{thab}, we see that 
a one-parameter family of Killing vectors 
gives one-parameter family of metrics $\tilde{\bs{h}}$ 
on the orbit space $\N$. 
Then, we should solve seven types of geodesic equations 
for finding all of cohomogeneity one Nambu-Goto string solutions 
in Minkowski spacetime. Some of them are easily solved
\cite{Frolov:1996xw}
\cite{Ogawa:etal}.

Generalizations of the present work to higher dimensional target spaces 
and higher dimensional cohomogeneity one objects are interesting issues. 
It is easy to get ordinary differential equations for 
higher dimensional cohomogeneity one objects, but 
classification of them would require some labors.


\begin{acknowledgments}
 We are grateful to Dr. K.~Nakao and Dr. Y.~Yasui for helpful discussions.
This work is supported by the Grant-in-Aid
for Scientific Research No.14540275. 
\end{acknowledgments}


\end{document}